# Thermally Oxidized Two-dimensional TaS$_2$ as a High-κ Gate Dielectric for MoS$_2$ Field-Effect Transistors

Bhim Chamlagain[1,#], Qingsong Cui[2,#], Sagar Paudel[1], Mark Ming-Cheng Cheng[2], Pai-Yen Chen[2], and Zhixian Zhou[1,*]

[1]Department of Physics and Astronomy, Wayne State University, Detroit, MI 48201, USA

[2]Department of Electrical and Computer Engineering, Wayne State University,

Detroit, MI 48202, USA

## Abstract

We report a new approach to integrating high-κ dielectrics in both bottom- and top-gated MoS$_2$ field-effect transistors (FETs) through thermal oxidation and mechanical assembly of layered two-dimensional (2D) TaS$_2$. Combined X-ray photoelectron spectroscopy (XPS), optical microscopy, atomic force microscopy (AFM), and capacitance-voltage (*C-V*) measurements confirm that multilayer TaS$_2$ flakes can be uniformly transformed to Ta$_2$O$_5$ with a high dielectric constant of ~ 15.5 via thermal oxidation, while preserving the geometry and ultra-smooth surfaces of 2D TMDs. Top-gated MoS$_2$ FETs fabricated using the thermally oxidized Ta$_2$O$_5$ as gate dielectric demonstrate a high current on/off ratio approaching 10$^6$, a subthreshold swing (SS) down to 61 mV/dec, and a field-effect mobility exceeding 60 cm$^2$V$^{-1}$ s$^{-1}$ at room temperature, indicating high dielectric quality and low interface trap density.

**Key words:** two-dimensional, MoS$_2$, TaS$_2$, Ta$_2$O$_5$, high-κ dielectric, thermal oxidation

[#] These authors contributed equally.

[*] Email of the corresponding author: zxzhou@wayne.edu

**Introduction**

Layered transition metal dichalcogenides (TMDs) such as $MoS_2$, $MoSe_2$, and $WSe_2$ have recently emerged as promising post-silicon electronic materials because they have not only demonstrated a multitude of graphene-like properties desirable for flexible electronics including a relatively high carrier mobility, mechanical flexibility, and chemical and thermal stability, but also offer the significant advantage of a substantial band gap essential for low-power digital electronics.[1-6] In addition, pristine surfaces of TMDs are free of dangling bonds, which reduces surface roughness scattering and interface traps. Recent experimental and theoretical studies have shown that the mobility of monolayer and multilayer TMDs such as $MoS_2$ and $MoSe_2$ is strongly affected by their dielectric environment and the quality of the interface between the channel and dielectric/substrate.[7-11] Ultraclean hexagonal boron nitride (h-BN) is an ideal substrate/dielectric material in preserving the intrinsic mobility of $MoS_2$ because it has atomically flat surfaces absent of dangling bonds, and is nearly free of charged impurities and charge traps.[12, 13] However, h-BN has a relative low dielectric constant of 3-4, while high-κ dielectrics are needed to optimize the electrostatic control of the channel and minimize operation voltage.[14-16]

In silicon-based electronics, atomic layer deposition (ALD) has been widely used to integrate high-κ gate dielectrics such as $Al_2O_3$ and $HfO_2$ with atomically controlled thickness and uniformity. However, the lack of out-of-plane covalent bonds or functional groups on pristine TMD surfaces imposes a significant challenge for high-κ dielectric integration in top-gated TMD devices because ALD requires chemical groups such as hydroxyl groups on the channel surfaces to form conformal and uniform interface layers.[17-20] To overcome this challenge, Liu *et al.* have significantly lowered the ALD temperature to grow $Al_2O_3$ films on $MoS_2$.[21] However, the coverage and uniformity of $Al_2O_3$ thin films grown by low temperature ALD are intricately affected by multiple growth parameters such as pulsing and purging times, process pressure, and cleanliness of the $MoS_2$ surface, leading to poor reproducibility.[22] For instance, small amount of organic and solvent residues on the TMD surfaces have been found to drastically influence the ALD nuclear behavior during the initial stage of deposition.[23] In addition, ALD dielectric films grown at low temperatures tend to contain substantial amount of impurities such as OH and C residues.[24, 25] A variety of surface functionalization methods such as oxygen plasma and ozone have been used to improve the smoothness and uniformity of ALD grown high-κ dielectrics on TMDs.[22, 26, 27] However, highly reactive oxygen plasma and ozone tend to degrade the electrical properties of monolayer and few-layer TMDs through surface oxidation and introduction of defect states. To date, it remains a major challenge to grow highly uniform, atomically smooth and ultrathin high-κ dielectrics on TMD surfaces while preserving the intrinsic channel properties of pristine TMD channels.

In this article, we present a new strategy to integrate high-κ dielectric into $MoS_2$ field-effect transistors through mechanical assembly of $Ta_2O_5$ chemically transformed from 2D $TaS_2$. In contrast to relatively inert semiconducting TMDs such as $MoS_2$, $MoSe_2$ and $WSe_2$, metallic TMDs such as $TaS_2$ are prone to surface oxidation in ambient environment.[28-30] At elevated temperatures, monolayer and multilayer $TaS_2$ can be chemically transformed into atomically flat, spatially uniform and nearly defect-free $Ta_2O_5$ insulator via thermal oxidation, as recently demonstrated by the authors.[31] In this work, we have systematically characterized the dielectric properties of $Ta_2O_5$ thermally oxidized from $TaS_2$ by capacitance-voltage measurement, yielding a thickness independent high dielectric constant of κ ~ 15.5. $MoS_2$ FETs fabricated using thermally oxidized thin $Ta_2O_5$ as gate dielectric show nearly hysteresis-free transfer characteristics, suggesting high interface quality. Furthermore, this new approach enables us to assemble high quality $Ta_2O_5$ dielectric on top of pristine MoS$_2$ channels to form top-gated FETs, while circumventing the constraints of ALD methods. MoS$_2$ FETs with the thermal Ta$_2$O$_5$ top-gate dielectric demonstrate a current on/off ratio of ~ 10$^6$, a high field-effect mobility > 60 cm$^2$V$^{-1}$s$^{-1}$, and a near ideal subthreshold swing (SS) of ~ 61 mV/dec at room temperature, along with pronounced drain-current saturation in their output characteristics. The impressive performance of the MoS$_2$ FETs can be attributed to the atomically flat, spatially uniform, and nearly charge-trap free high-κ Ta$_2$O$_5$ dielectric synthesized by thermal oxidation of TaS$_2$, which forms ultraclean interface with the MoS$_2$ channel.

**Results and discussion**

To chemically transform TaS$_2$ 2D metal into Ta$_2$O$_5$ high-κ dielectric, TaS$_2$ flakes were first mechanically exfoliated on SiO$_2$/Si substrate and subsequently oxidized by heating on a hot plate at 300°C for 3 hours in air. The chemical transformation of TaS$_2$ into Ta$_2$O$_5$ was confirmed by X-ray photoelectron spectroscopy (XPS). **Figure 1a** shows Ta 4f core level XPS spectra of multilayer TaS$_2$ flakes that are (i) freshly exfoliated (inserted into the XPS chamber within 1-2 minutes of exfoliation), (ii) exposed in ambient environment for one day, and (iii) heated at 300°C in ambient air for 3 hrs. The freshly exfoliated TaS$_2$ flakes exhibit two well-defined peaks at binding energies of 22.8 eV and 24.7 eV, corresponding to the Ta$^{4+}$ 4f$_{7/2}$ and Ta$^{4+}$ 4f$_{5/2}$ doublet of TaS$_2$. After exposed to ambient air for a day, the TaS$_2$ flakes exhibit two additional weak peaks at slightly higher binding energies, suggesting partial oxidation of TaS$_2$ surfaces. After the TaS$_2$ flakes were heated to 300°C for 3 hours in air, the Ta$^{4+}$ doublelet of TaS$_2$ completely disappears, while the weak peaks develop into two distinct peaks at binding energies of 25.7 eV and 27.6 eV, corresponding to the Ta$^{5+}$ 4f$_{7/2}$ and Ta$^{5+}$ 4f$_{5/2}$ doublet of Ta$_2$O$_5$.[32-35] The chemical transformation of TaS$_2$ into $Ta_2O_5$ is also clearly manifested in the form of color change. The lower-right and upper-left insets of **Figure 1b** present the optical micrographs of a ~ 12 nm multilayer $TaS_2$ sample before and after thermal oxidation (which chemically transforms $TaS_2$ to $Ta_2O_5$). While the sample geometry and lateral dimensions remain nearly unchanged throughout the chemical transformation, the sample thickness measured by AFM systematically decreases by ~4% upon thermal oxidation independent of initial sample thickness as shown in **Figure 1b**, indicating that thermal oxidation occurs throughout the entire sample. Because about 25% volume decrease would be expected if the $TaS_2$ was transformed into crystalline $Ta_2O_5$, the 4% thickness reduction (with nearly unchanged lateral dimensions) observed here suggests that the $Ta_2O_5$ has significantly lower density than crystalline $Ta_2O_5$ and is likely amorphous.[36, 37] The non-crystalline structure of our $Ta_2O_5$ with relatively low density allows the oxygen in air to diffuse deep into the sample and achieve uniform and thorough thermal oxidation. In spite of the chemical and structural transformation, the $Ta_2O_5$ flakes display very low root-mean-square (RMS) surface roughness. **Figure 1c** shows a representative AFM topographic image acquired on a ~ 12 nm thick $Ta_2O_5$ synthesized by thermal oxidation yielding a RMS roughness of 0.17 nm, which is comparable to that of TMDs.[38] Such a low RMS roughness in the thermal $Ta_2O_5$ is essential for the formation of high quality interface with $MoS_2$ and other TMD channel materials.

To extract the dielectric constant of the $Ta_2O_5$ synthesized by thermal oxidation, we fabricated capacitors consisting of an ultrathin $Ta_2O_5$ dielectric layer sandwiched between top and bottom metal electrodes as schematically illustrated in **Figure 2a** (See the **Methods**). Briefly, bottom electrodes consisting of 5 nm of titanium (Ti) adhesion layer and 10 nm of platinum (Pt) were first fabricated on $Si/SiO_2$ substrate. Next, selected ultrathin $TaS_2$ flakes produced by mechanical exfoliation were placed on top of the Pt bottom electrodes using a dry transfer method.[31, 39] Subsequently, the TaS$_2$ flakes on Pt were thermally oxidized to Ta$_2$O$_5$ in air. Finally, multiple top electrodes consisting of 10 nm Ti and 30 nm Au were fabricated on top of the Ta$_2$O$_5$ dielectric, where the area of each metal-insulator-metal capacitor is defined by the area of the Ta$_2$O$_5$ dielectric in the overlap region between each top electrode and the bottom electrode. **Figure 2b** shows the capacitance as a function of the applied DC bias voltage (*C-V*) of a capacitor with a 13 nm thick Ta$_2$O$_5$ dielectric measured by applying a 500 Hz and 50 mV AC excitation voltage. The nearly voltage independent capacitance shows negligible hysteresis, indicating that the Ta$_2$O$_5$ dielectric is of high quality because charge traps in the dielectric and at the interfaces usually introduce non-negligible hysteresis. To exclude the background capacitance (e.g. the cable capacitance), we plot the total capacitance as a function of area for capacitors fabricated using the same piece of Ta$_2$O$_5$ with uniform thickness, as shown in **figure 2c**. From a linear fit to the capacitance *vs*. area data, capacitance per unit area (*C/A*) of the capacitors with different Ta$_2$O$_5$ dielectric thicknesses is determined. The near zero intercept of the linear fit in **figure 2c** indicates negligible background capacitance. The dielectric constant can be calculated using the parallel capacitor model: $\frac{C}{A} = \frac{\varepsilon_0 \kappa}{t}$, where $\varepsilon_0$ is the permittivity of the free space, and $\kappa$ and $t$ are the dielectric constant and thickness of the Ta$_2$O$_5$, respectively. **Figure 2d** shows that the inverse of capacitance per unit area (*A/C*) is linearly proportional to the thickness (*t*) for various Ta$_2$O$_5$ thicknesses ranging from 5 to 33 nm, indicating that the dielectric constant of the Ta$_2$O$_5$ is thickness independent. This finding provides further evidence that multilayer TaS$_2$ samples have been uniformly transformed into Ta$_2$O$_5$. From the slope of the linear fit, we extract a dielectric constant of ~ 15.5, which is consistent with the $\kappa$ reported for amorphous Ta$_2$O$_5$.[40] It is worth noting that the dielectric constant of our thermal Ta$_2$O$_5$ is about 4 times larger than that of SiO$_2$ (3.9) and h-BN (3 - 4).

To further evaluate the quality of the thermally oxidized $Ta_2O_5$ as a high-κ dielectric for 2D electronics, we first fabricated $MoS_2$ FETs with an ultrathin $Ta_2O_5$ dielectric and a multilayer graphene (M-Gr) gate as schematically shown in **Figure 3a**. Here we choose multiplayer graphene as the gate because it not only has atomically smooth surfaces but also is compatible with the fabrication of 2D flexible electronics in the future. To fabricate the devices, mechanically exfoliated multilayer $TaS_2$ flakes were dry-transferred on top of mechanically exfoliated M-Gr flakes serving as gate electrodes, and subsequently thermally oxidized to $Ta_2O_5$. Next, a selected few-layer $MoS_2$ channel was placed on top of the $Ta_2O_5$/M-Gr stack also by dry-transfer.[39] Finally, drain, source and gate contacts were fabricated by electron beam lithography and subsequent deposition of 5 nm Ti and 40 nm Au (see the **Methods**). **Figure 3b** presents a micrograph of a representative $MoS_2$ FET consisting of a 7.0 nm thick $MoS_2$ channel, a 6.5 nm thick $Ta_2O_5$ dielectric and a multilayer graphene back gate. **Figure 3c** shows the output characteristics of the $MoS_2$ device depicted in **Figure 3b**. In the low $V_{ds}$ region, the *I-V* characteristics are linear, indicating near ohmic contacts. At high drain/source voltages, the device exhibits apparent current saturation partially due to the channel pinch off, suggesting effective gate coupling. **Figure 3d** presents room-temperature transfer characteristics of the device measured at $V_{ds}$ = 100 mV. The $MoS_2$ device exhibits *n*-type behavior with a current on/off ratio exceeding $10^5$, where the off current is limited by the leak current as shown in **Figure 3d**. In spite of the ultrathin layer thickness (6.5 nm) and relatively small band bap of $Ta_2O_5$ (3.8-5.3 eV) as a dielectric, the gate leak current is rather low, suggesting that the thermally oxidized $Ta_2O_5$ is highly uniform with very low density of pinholes. We expect that the gate-leak current can be significantly reduced by using high-κ dielectrics with a larger bandgap such as $HfO_2$. Similar to metallic $TaS_2$, $HfSe_2$ is also prone to oxidation in air in spite of the fact that it is a semiconductor with a bulk band-gap comparable to that of balk MoS$_2$.[41, 42] We envision that 2D HfSe$_2$ can be relatively straightforwardly transformed to ultrathin HfO$_2$ high-κ dielectric and integrated into FETs with semiconducting TMD channel materials such as MoS$_2$ by mechanical assembly. The on-current can be increased through contact engineering to further improve the on/off ratio and overall device performance.[39] The transfer characteristics measured with opposite gate sweep directions show negligible hysteresis, indicating low charge trap density at the channel/dielectric interface. The subthreshold swing of the device (~ 64 mV/dec) approaches the theoretical limit of $\frac{kT}{e}\ln(10) = 60\ mV/dec$ at $T$ = 300 K, which can be attributed to the large gate capacitance of the ultrathin Ta$_2$O$_5$ high-κ dielectric and high interface quality. A relatively low interface trap density of $D_{it}$ = 1.2×10$^{12}$ cm$^{-2}$eV$^{-1}$ is calculated from the following equations:

$$SS = (1 + \frac{C_{it}}{C_{ox}}) \times kT \times ln(10)/e \quad (1)$$

$$D_{it} = C_{it}/e \quad (2)$$

Here $C_{it}$ and $C_{ox}$ are the interfacial and oxide capacitances, respectively.

For practical application of MoS$_2$ as a channel material in integrated circuits, top-gated MoS$_2$ FETs with high-κ gate dielectric are needed to individually control each device. A significant advantage of this approach to integrating high-κ dielectrics in 2D electronics is that it circumvents a major challenge encountered in the ALD growth of high-κ dielectrics on TMDs due to the lack of dangling bonds on semiconducting TMD surfaces. **Figure 4a** schematically shows a few-layer MoS$_2$ FET with Ta$_2$O$_5$ as top-gate dielectric. To fabricate the devices, ultrathin flakes of Ta$_2$O$_5$ were produced by thermal oxidation of mechanically exfoliated TaS$_2$ multilayers on Si substrate at 300°C in air, and subsequently transferred onto mechanically exfoliated few-layer MoS$_2$ on degenerately doped Si substrate with 290 nm thermal oxide. Both the drain/source and top-gate electrodes were then fabricated by EBL and e-beam deposition of 5 nm Ti and 40 nm Au. **Figure**

**4b** shows an optical micrograph of a representative top-gated MoS$_2$ FET, which consists of a 6.5 nm thick MoS$_2$ channel and 31 nm thick Ta$_2$O$_5$ gate dielectric. As shown in **Figure 4c**, the output characteristics of the device show linear behavior at low $V_{ds}$ and current saturation at high $V_{ds}$, similar to the MoS$_2$ devices with Ta$_2$O$_5$ bottom-gate dielectric. Here a constant back-gate voltage of 60 V is applied to reduce the contact resistance and turn on the under-lapped regions between the drain/source electrodes and top-gate electrode. The higher on-current observed in the top-gated MoS$_2$ device than in the bottom-gated MoS$_2$ device shown in **Figure 3** can be chiefly attributed to the reduced contact resistance.

**Figure 4d** shows room temperature transfer characteristics of the same device measured at $V_{ds}$ = 100 mV by sweeping the top-gate voltage at a fixed back-gate voltage of 60 V. The transfer curve exhibits a nearly ideal SS of 61 mV/dec in spite of the relatively thick (31 nm) Ta$_2$O$_5$ gate dielectric, indicating nearly trap-free channel/dielectric interface. The current on/off ratio approaches $10^6$, which can be further enhanced by increasing the on-current through the reduction of series resistance. It is worth noting that the drain current starts to saturate at $V_{tg} \sim 0$ V, which is mainly caused by the reduction of the effective gate voltage ($V_{tg\_eff}$) due to the presence of a significant series resistance from the drain/source contacts and the under-lapped regions given by $V_{tg\_eff} = V_{tg} - R_s I_{ds}$, where $V_{tg}$ is the applied top-gate voltage, and $R_s$ is the sum of metal/MoS$_2$ contact resistance and the resistance of the under-lapped regions.[43] The presence of a substantial $R_s$ may also partially contribute to the current saturation in the output characteristics because the effective drain-source bias voltage given by $V_{ds\_eff} = V_{ds} - 2R_s I_{ds}$ is also reduced at high $I_{ds}$. As shown in **Figure 4d**, a field-effect mobility of $\mu_{FE} \approx 61.5$ cm$^2$V$^{-1}$s$^{-1}$ is extracted from the linear region of the transfer curve using the expression $\mu_{FE} = \frac{L}{W} \times \frac{dI_{ds}}{dV_{tg}} \times \frac{1}{C_{tg}} \times \frac{1}{V_{ds}}$. Here $L$ is the channel length directly underneath the metal gate because the under-lapped regions are not tunable by the top-gate. Inclusion of the under-lapped areas in the calculations would overestimate the field-effect mobility. The field-effect mobility observed in our top-gated MoS$_2$ FETs is comparable to the highest room-temperature mobility values for top-gated MoS$_2$ devices with high-κ dielectric, further indicating low density of trap states in the thermally oxidized Ta$_2$O$_5$ dielectric and at the channel/dielectric interface. Furthermore, the reduced trap density in conjunction with the relatively high dielectric constant of the thermally oxidized Ta$_2$O$_5$ facilitates a steeper subthreshold slope (SS ≈ 61 mV/dec) than that previously reported on MoS$_2$ devices with hBN and high-κ dielectrics (SS > 70 mV/dec). [14, 44-46]

**Conclusion**

In summary, we have demonstrated integration of ultrathin Ta$_2$O$_5$ chemically transformed from 2D TaS$_2$ in both bottom-gated and top-gated MoS$_2$ FETs as a high-κ gate dielectric. These devices show desirable FET characteristics such as a high on/off ratio, absence of hysteresis, a nearly ideal subthreshold swing, and a relatively high mobility, indicating high dielectric and interface quality. The newly developed dielectric integration strategy via chemical transformation of 2D materials to high-κ dielectrics in conjunction with polymer-based dry transfer techniques overcomes a significant challenge of dielectric integration in semiconducting TMD devices, and is applicable to a wide range of 2D materials. This approach is also scalable when combined with large area synthesis techniques such as chemical vapor deposition (CVD) and liquid exfoliation.[47, 48]

**Methods**

*XPS measurement*

X-ray photoelectron spectroscopy (XPS) measurement was performed using a Kratos Axis Ultra XPS system with a monochromatic Al source. Because TaS$_2$ is sensitive to air, the samples were mechanically exfoliated from bulk crystals right before XPS measurement and then immediately inserted into the XPS chamber to avoid oxidation unless specified otherwise in the manuscript. Pass energy of 20 eV with 0.1 eV scanning step was used for photoelectron detection. The C 1s reference line at the binding energy of 284.6 eV was used to calibrate the charging effect.

*Capacitance-voltage (C-V) measurement*

For *C-V* measurement, parallel-plate capacitors were fabricated by sandwiching an ultrathin Ta$_2$O$_5$ between a pair of metal electrodes. First, bottom electrodes consisting of 10 nm of platinum (Pt) with 5 nm of titanium (Ti) adhesion layer were patterned on Si substrates with 290 nm of thermal oxide using electron beam lithography followed by electron beam deposition and lift-off. Next, TaS$_2$ flakes were produced by mechanical exfoliation of commercial available TaS$_2$ crystals on Poly-dimethylsiloxane (PDMS) stamps and subsequently transferred on top of the Pt electrodes.[49] The TaS$_2$ flakes were then thermally oxidized to Ta$_2$O$_5$ by heating on a hotplate at 300°C for 3 hours in ambient air. Finally, top electrodes were fabricated by e-beam lithography and electron beam deposition of 10 nm Ti and 30 nm Au. The *C-V* measurements were carried out at room temperature using an Agilent 4284A Precision LCR Meter inside a Lakeshore TTPX probe station.

*MoS$_2$ FET fabrication and electrical measurements*

To fabricate MoS$_2$ FETs with Ta$_2$O$_5$ gate dielectric and graphite bottom gate, thin graphite flakes were first mechanically exfoliated and transferred onto Si/SiO$_2$ substrates as bottom gates. Subsequently, multilayer TaS$_2$ flakes were produced by mechanical exfoliation from commercially available TaS$_2$ crystals on PDMS stamps and subsequently transferred onto top of the graphite gates using a home-built precision transfer stage. The substrates were then heated at 300°C for 3 hrs in ambient environment to chemically transform $TaS_2$ to $Ta_2O_5$. The oxidized flakes were further characterized by optical microscope and XE-70 non-contact mode atomic force microscopy (AFM). Next, few layer $MoS_2$ flakes were exfoliated on PDMS stamp and transferred onto the $Ta_2O_5$/graphite gate stack. Finally, drain/source electrodes and electrical contacts to the graphite gate were fabricated by e-beam lithography and electron beam deposition of 5 nm of Ti and 40 nm of Au followed by acetone lift-off.

To fabricate top-gated $MoS_2$ FETs, mechanically exfoliated multilayer $TaS_2$ flakes were firstly thermally oxidized to $Ta_2O_5$ on $Si/SiO_2$ substrate. Next, the $Ta_2O_5$ flakes were picked up using polycarbonate (PC) and transferred onto the top of $MoS_2$ FETs pre-fabricated on $SiO_2/Si$ substrates. Finally, top gate electrodes were fabricated on top of the $Ta_2O_5$ gate dielectric by e-beam lithography and subsequent electron beam deposition of 10 nm of Ti covered by 40 nm of Au.

Transport properties of the fabricated $MoS_2$ FET devices were measured by a Keithley 4200-SCS semiconductor parameter analyzer in a Lakeshore TTPX probe station at room temperature and in high vacuum ($1\times10^{-6}$ Torr).

**Acknowledgement**

B.C. and Z.Z. acknowledge partial support by NSF grant number DMR-1308436 and the WSU Presidential Research Enhancement Award. P.Y.C. would like to acknowledge the Air Force Research Laboratory Summer Faculty Fellowship Program (ARFL-SFFP).

*Conflicts of interest:* The authors declare no competing financial interest.

# References

1. Wang, Q. H.; Kalantar-Zadeh, K.; Kis, A.; Coleman, J. N.; Strano, M. S. *Nat Nano* **2012,** 7, (11), 699-712.
2. Podzorov, V.; Gershenson, M. E.; Kloc, C.; Zeis, R.; Bucher, E. *Appl. Phys. Lett.* **2004,** 84, (17), 3301-3303.
3. Fang, H.; Chuang, S.; Chang, T. C.; Takei, K.; Takahashi, T.; Javey, A. *Nano Lett.* **2012,** 12, 3788–3792.
4. Fang, H.; Tosun, M.; Seol, G.; Chang, T. C.; Takei, K.; Guo, J.; Javey, A. *Nano Lett.* **2013,** 13, (5), 1991-1995.
5. Liu, W.; Kang, J.; Sarkar, D.; Khatami, Y.; Jena, D.; Banerjee, K. *Nano Lett.* **2013,** 13, (5), 1983-1990.
6. Chuang, H.-J.; Tan, X.; Ghimire, N. J.; Perera, M. M.; Chamlagain, B.; Cheng, M. M.-C.; Yan, J.; Mandrus, D.; Tománek, D.; Zhou, Z. *Nano Letters* **2014,** 14, (6), 3594-3601.
7. Bao, W.; Cai, X.; Kim, D.; Sridhara, K.; Fuhrer, M. S. *Appl. Phys. Lett.* **2013,** 102, (4), 042104.
8. Radisavljevic, B.; Kis, A. *Nat Mater* **2013,** 12, (9), 815-820.
9. Kim, S.; Konar, A.; Hwang, W.-S.; Lee, J. H.; Lee, J.; Yang, J.; Jung, C.; Kim, H.; Yoo, J.-B.; Choi, J.-Y.; Jin, Y. W.; Lee, S. Y.; Jena, D.; Choi, W.; Kim, K. *Nature Commun.* **2012,** 3, 1011.
10. Zeng, L.; Xin, Z.; Chen, S.; Du, G.; Kang, J.; Liu, X. *Appl. Phys. Lett.* **2013,** 103, (11), 113505.
11. Chamlagain, B.; Li, Q.; Ghimire, N. J.; Chuang, H.-J.; Perera, M. M.; Tu, H.; Xu, Y.; Pan, M.; Xaio, D.; Yan, J.; Mandrus, D.; Zhou, Z. *ACS Nano* **2014,** 8, (5), 5079-5088.
12. Dean, C. R.; Young, A. F.; MericI; LeeC; WangL; SorgenfreiS; WatanabeK; TaniguchiT; KimP; Shepard, K. L.; HoneJ. *Nat Nano* **2010,** 5, (10), 722-726.
13. Cui, X.; Lee, G.-H.; Kim, Y. D.; Arefe, G.; Huang, P. Y.; Lee, C.-H.; Chenet, D. A.; Zhang, X.; Wang, L.; Ye, F.; Pizzocchero, F.; Jessen, B. S.; Watanabe, K.; Taniguchi, T.; Muller, D. A.; Low, T.; Kim, P.; Hone, J. *Nat Nano* **2015,** 10, (6), 534-540.
14. Lee, G.-H.; Yu, Y.-J.; Cui, X.; Petrone, N.; Lee, C.-H.; Choi, M. S.; Lee, D.-Y.; Lee, C.; Yoo, W. J.; Watanabe, K.; Taniguchi, T.; Nuckolls, C.; Kim, P.; Hone, J. *ACS Nano* **2013,** 7, (9), 7931-7936.
15. Lembke, D.; Bertolazzi, S.; Kis, A. *Acc Chem Res* **2015,** 48, (1), 100-10.
16. Kim, S.; Konar, A.; Hwang, W. S.; Lee, J. H.; Lee, J.; Yang, J.; Jung, C.; Kim, H.; Yoo, J. B.; Choi, J. Y.; Jin, Y. W.; Lee, S. Y.; Jena, D.; Choi, W.; Kim, K. *Nat Commun* **2012,** 3, 1011.
17. Kim, S. K.; Lee, S. W.; Hwang, C. S.; Min, Y.-S.; Won, J. Y.; Jeong, J. *Journal of The Electrochemical Society* **2006,** 153, (5), F69.
18. Azcatl, A.; McDonnell, S.; K. C, S.; Peng, X.; Dong, H.; Qin, X.; Addou, R.; Mordi, G. I.; Lu, N.; Kim, J.; Kim, M. J.; Cho, K.; Wallace, R. M. *Applied Physics Letters* **2014,** 104, (11), 111601.
19. Cheng, L.; Qin, X.; Lucero, A. T.; Azcatl, A.; Huang, J.; Wallace, R. M.; Cho, K.; Kim, J. *ACS Appl Mater Interfaces* **2014,** 6, (15), 11834-8.
20. Yang, J.; Kim, S.; Choi, W.; Park, S. H.; Jung, Y.; Cho, M. H.; Kim, H. *ACS Appl Mater Interfaces* **2013,** 5, (11), 4739-44.
21. Liu, H.; Xu, K.; Zhang, X.; Ye, P. D. *Appl. Phys. Lett.* **2012,** 100, (15), 152115.
22. Cheng, L.; Qin, X.; Lucero, A. T.; Azcatl, A.; Huang, J.; Wallace, R. M.; Cho, K.; Kim, J. *ACS Applied Materials & Interfaces* **2014,** 6, (15), 11834-11838.
23. McDonnell, S.; Brennan, B.; Azcatl, A.; Lu, N.; Dong, H.; Buie, C.; Kim, J.; Hinkle, C. L.; Kim, M. J.; Wallace, R. M. *ACS Nano* **2013,** 7, (11), 10354-10361.
24. Kim, S. K.; Lee, S. W.; Hwang, C. S.; Min, Y.-S.; Won, J. Y.; Jeong, J. *Journal of the Electrochemical Society* **2006,** 153, (5), F69-F76.
25. Wirtz, C.; Hallam, T.; Cullen, C. P.; Berner, N. C.; O'Brien, M.; Marcia, M.; Hirsch, A.; Duesberg, G. S. *Chem Commun (Camb)* **2015,** 51, (92), 16553-6.

26. Azcatl, A.; McDonnell, S.; K. C., S.; Peng, X.; Dong, H.; Qin, X.; Addou, R.; Mordi, G. I.; Lu, N.; Kim, J.; Kim, M. J.; Cho, K.; Wallace, R. M. *Applied Physics Letters* **2014,** 104, (11), 111601.
27. Yang, J.; Kim, S.; Choi, W.; Park, S. H.; Jung, Y.; Cho, M.-H.; Kim, H. *ACS Applied Materials & Interfaces* **2013,** 5, (11), 4739-4744.
28. Geim, A. K.; Grigorieva, I. V. *Nature* **2013,** 499, (7459), 419-425.
29. Cao, Y.; Mishchenko, A.; Yu, G. L.; Khestanova, E.; Rooney, A. P.; Prestat, E.; Kretinin, A. V.; Blake, P.; Shalom, M. B.; Woods, C.; Chapman, J.; Balakrishnan, G.; Grigorieva, I. V.; Novoselov, K. S.; Piot, B. A.; Potemski, M.; Watanabe, K.; Taniguchi, T.; Haigh, S. J.; Geim, A. K.; Gorbachev, R. V. *Nano Letters* **2015,** 15, (8), 4914-4921.
30. Tsen, A. W.; Hovden, R.; Wang, D.; Kim, Y. D.; Okamoto, J.; Spoth, K. A.; Liu, Y.; Lu, W.; Sun, Y.; Hone, J. C.; Kourkoutis, L. F.; Kim, P.; Pasupathy, A. N. *Proceedings of the National Academy of Sciences* **2015,** 112, (49), 15054-15059.
31. Cui, Q.; Sakhdari, M.; Chamlagain, B.; Chuang, H.-J.; Liu, Y.; Cheng, M. M.-C.; Zhou, Z.; Chen, P.-Y. *ACS Applied Materials & Interfaces* **2016**.
32. Martinez, H.; Auriel, C.; Gonbeau, D.; Loudet, M.; PfisterGuillouzo, G. *Applied Surface Science* **1996,** 93, (3), 231-235.
33. Aminpirooz, S.; Becker, L.; Rossner, H.; Schellenberger, A.; Holub-Krappe, E. *Surface Science* **1995,** 331–333, Part A, 501-505.
34. Atanassova, E.; Tyuliev, G.; Paskaleva, A.; Spassov, D.; Kostov, K. *Applied Surface Science* **2004,** 225, (1-4), 86-99.
35. Ho, S. F.; Contarini, S.; Rabalais, J. W. *The Journal of Physical Chemistry* **1987,** 91, (18), 4779-4788.
36. Revelli, J. F.; Disalvo, F. J., Tantalum Disulfide (TaS2) and Its Intercalation Compounds. In *Inorganic Syntheses*, John Wiley & Sons, Inc.: 2007; pp 155-169.
37. Reisman, A.; Holtzberg, F.; Berkenblit, M.; Berry, M. *Journal of the American Chemical Society* **1956,** 78, (18), 4514-4520.
38. Taeg Yeoung, K.; Areum, J.; Wontaek, K.; Jinhwan, L.; Youngchan, K.; Jung Eun, L.; Gyeong Hee, R.; Kwanghee, P.; Dogyeong, K.; Zonghoon, L.; Min Hyung, L.; Changgu, L.; Sunmin, R. *2D Materials* **2017,** 4, (1), 014003.
39. Chuang, H.-J.; Chamlagain, B.; Koehler, M.; Perera, M. M.; Yan, J.; Mandrus, D.; Tománek, D.; Zhou, Z. *Nano Letters* **2016,** 16, (3), 1896-1902.
40. Chang, C.-S.; Liu, T.-P.; Wu, T.-B. *Journal of Applied Physics* **2000,** 88, (12), 7242-7248.
41. Yue, R.; Barton, A. T.; Zhu, H.; Azcatl, A.; Pena, L. F.; Wang, J.; Peng, X.; Lu, N.; Cheng, L.; Addou, R.; McDonnell, S.; Colombo, L.; Hsu, J. W. P.; Kim, J.; Kim, M. J.; Wallace, R. M.; Hinkle, C. L. *ACS Nano* **2015,** 9, (1), 474-480.
42. Mleczko, M. J.; Zhang, C.; Lee, H. R.; Kuo, H. H.; Magyari-Köpe, B.; Shen, Z. X.; Moore, R. G.; Fisher, I. R.; Nishi, Y.; Pop, E. In *Atomically-thin HfSe$_2$ transistors with native metal oxides*, 2016 74th Annual Device Research Conference (DRC), 19-22 June 2016, 2016; 2016; pp 1-2.
43. Liu, W.; Kang, J.; Sarkar, D.; Khatami, Y.; Jena, D.; Banerjee, K. *Nano Lett* **2013,** 13, (5), 1983-90.
44. Lee, G.-H.; Cui, X.; Kim, Y. D.; Arefe, G.; Zhang, X.; Lee, C.-H.; Ye, F.; Watanabe, K.; Taniguchi, T.; Kim, P.; Hone, J. *ACS Nano* **2015,** 9, (7), 7019-7026.
45. Radisavljevic, B.; Radenovic, A.; Brivio, J.; Giacometti, V.; Kis, A. *Nature Nanotech.* **2011,** 6, (3), 147-150.
46. Yang, W.; Sun, Q.-Q.; Geng, Y.; Chen, L.; Zhou, P.; Ding, S.-J.; Zhang, D. W. *Scientific Reports* **2015,** 5, 11921.
47. Huang, J.-K.; Pu, J.; Hsu, C.-L.; Chiu, M.-H.; Juang, Z.-Y.; Chang, Y.-H.; Chang, W.-H.; Iwasa, Y.; Takenobu, T.; Li, L.-J. *ACS Nano* **2013,** 8, (1), 923-930.

48. Zhu, J.; Kang, J.; Kang, J.; Jariwala, D.; Wood, J. D.; Seo, J.-W. T.; Chen, K.-S.; Marks, T. J.; Hersam, M. C. *Nano Letters* **2015,** 15, (10), 7029-7036.
49. Li, H.; Wu, J.; Huang, X.; Yin, Z.; Liu, J.; Zhang, H. *ACS Nano* **2014,** 8, (7), 6563-70.

**Figure Captions**

**Figure 1. (a)** XPS of mechanically exfoliated TaS$_2$ flakes measured within 2 min of ambient exposure (i), after 1 day of ambient exposure (ii), and after 3hr of thermal oxidation at 300°C in air (iii). **(b)** Thicknesses of mechanically exfoliated TaS$_2$ flakes before and after 3hr of thermal oxidation at 300°C in air as determined by AFM. The insets show optical micrographs of a typical TaS$_2$ flake on SiO$_2$ before (bottom right) and after (top left) thermal oxidation at 300°C in air. **(c)** AFM surface topography of a Ta$_2$O$_5$ flake converted from a corresponding TaS$_2$.

**Figure 2. (a)** Schematic illustration of capacitors consisting of a thin Ta$_2$O$_5$ dielectric sandwiched between Pt and Ti/Au metal electrodes (b) Capacitance *vs.* voltage for a capacitor comprising a 13 nm thick Ta$_2$O$_5$ dielectric as the voltage is swept along both negative and positive directions. (c) Capacitance as a function of area for three capacitors with the same Ta$_2$O$_5$ dielectric thickness (13 nm) measured at two different frequencies (500 Hz and 1 kHz). Inset: an optical image of the corresponding capacitors. (d) A/C *vs.* thickness of Ta$_2$O$_5$. The inverse of slope of the plot gives the dielectric constant of Ta$_2$O$_5$ based on the parallel plate capacitor model: $\frac{C}{A} = \frac{\varepsilon_0 \kappa}{t}$.

**Figure 3.** (a) Schematic illustration of a MoS$_2$ FET device with Ta$_2$O$_5$ dielectric and multilayer-graphene (M-Gr) bottom gate. (b) Optical image of a typical bottom-gated MoS$_2$ FET device with Ta$_2$O$_5$ dielectric. (c) Output characteristics of the MoS$_2$ device shown in (b). (d) Transfer characteristics of the same MoS$_2$ FET device along with the gate leakage current. Red color represents the positive sweep direction and the blue color represents the negative sweep direction of the gate voltage.

**Figure 4.** (a) Cross-sectional view of a MoS$_2$ FET device with Ta$_2$O$_5$ dielectric and metal top gate. (b) Optical image of a MoS$_2$ FET device with Ta$_2$O$_5$ top-gate dielectric. **(c)** Output characteristics of the MoS$_2$ device in (b). (d) Transfer characteristics of the same MoS$_2$ FET device plotted both in semi-log and linear scales.

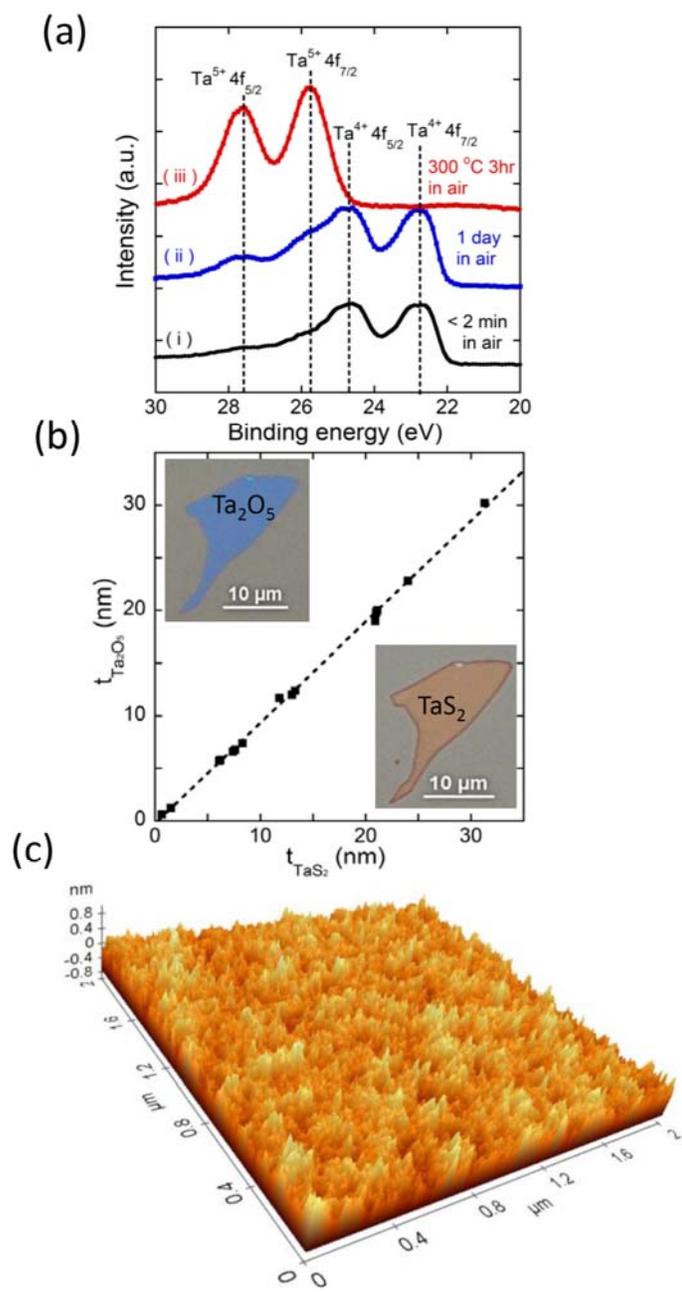

Figure 1

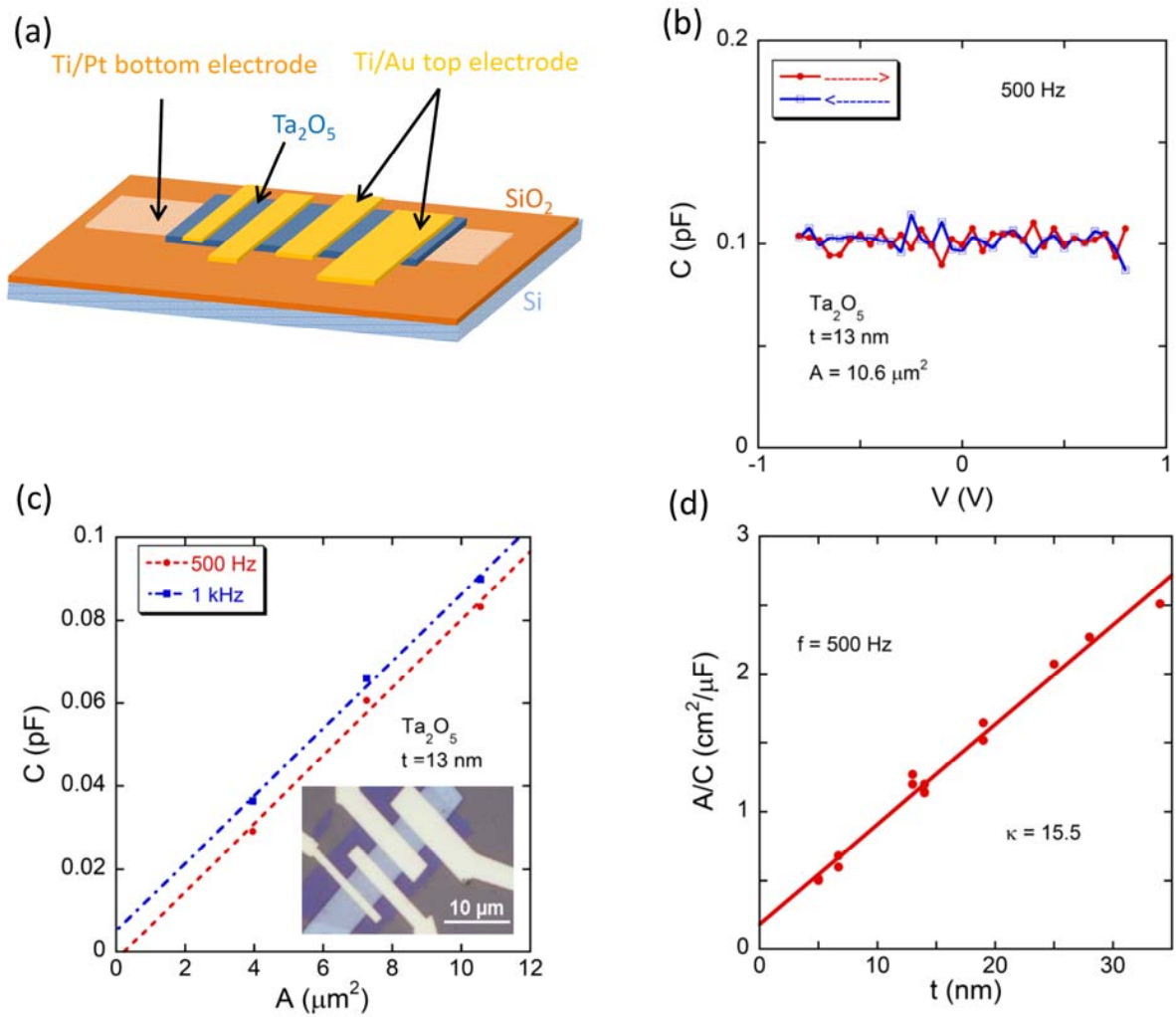

Figure 2

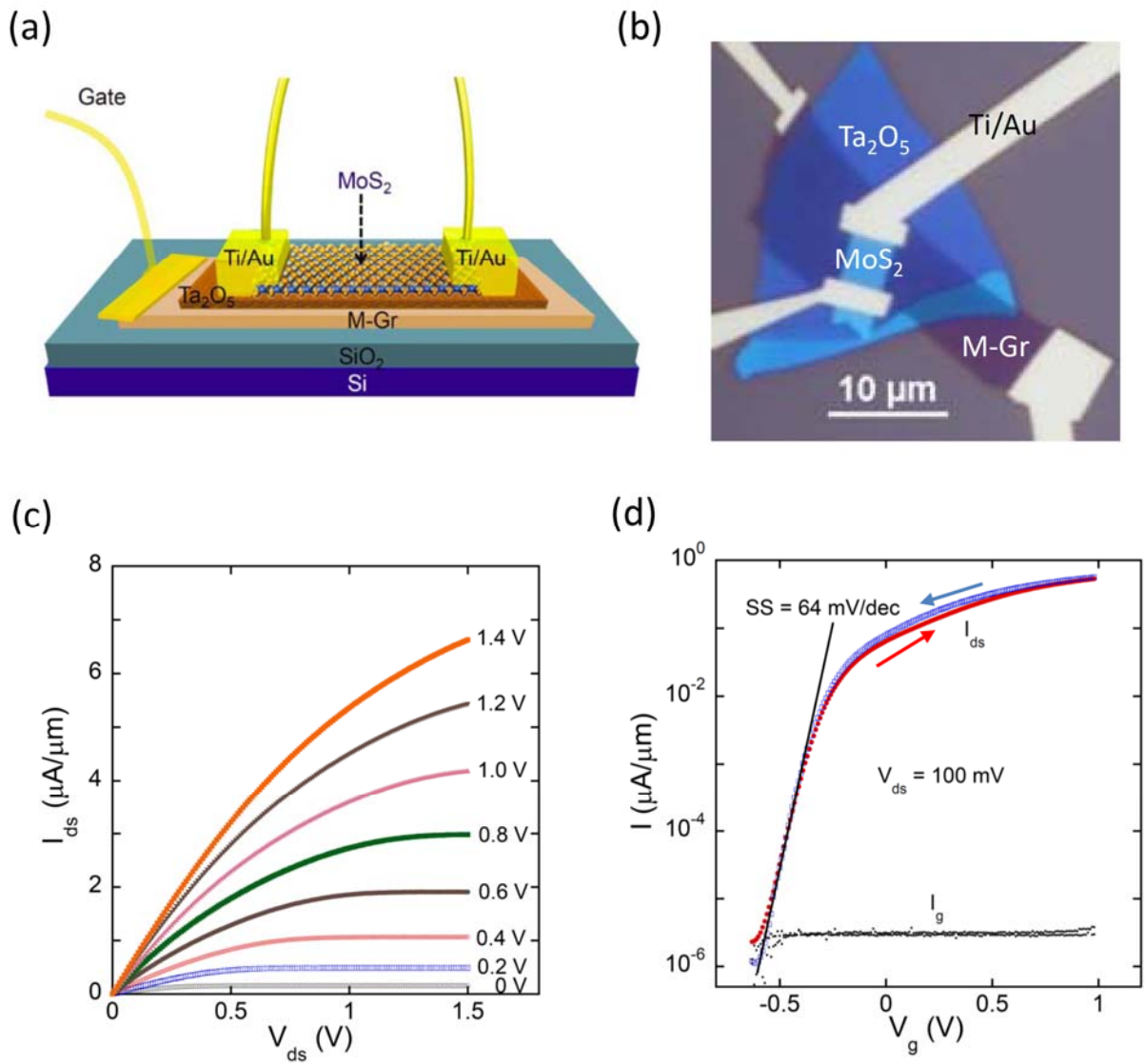

Figure 3

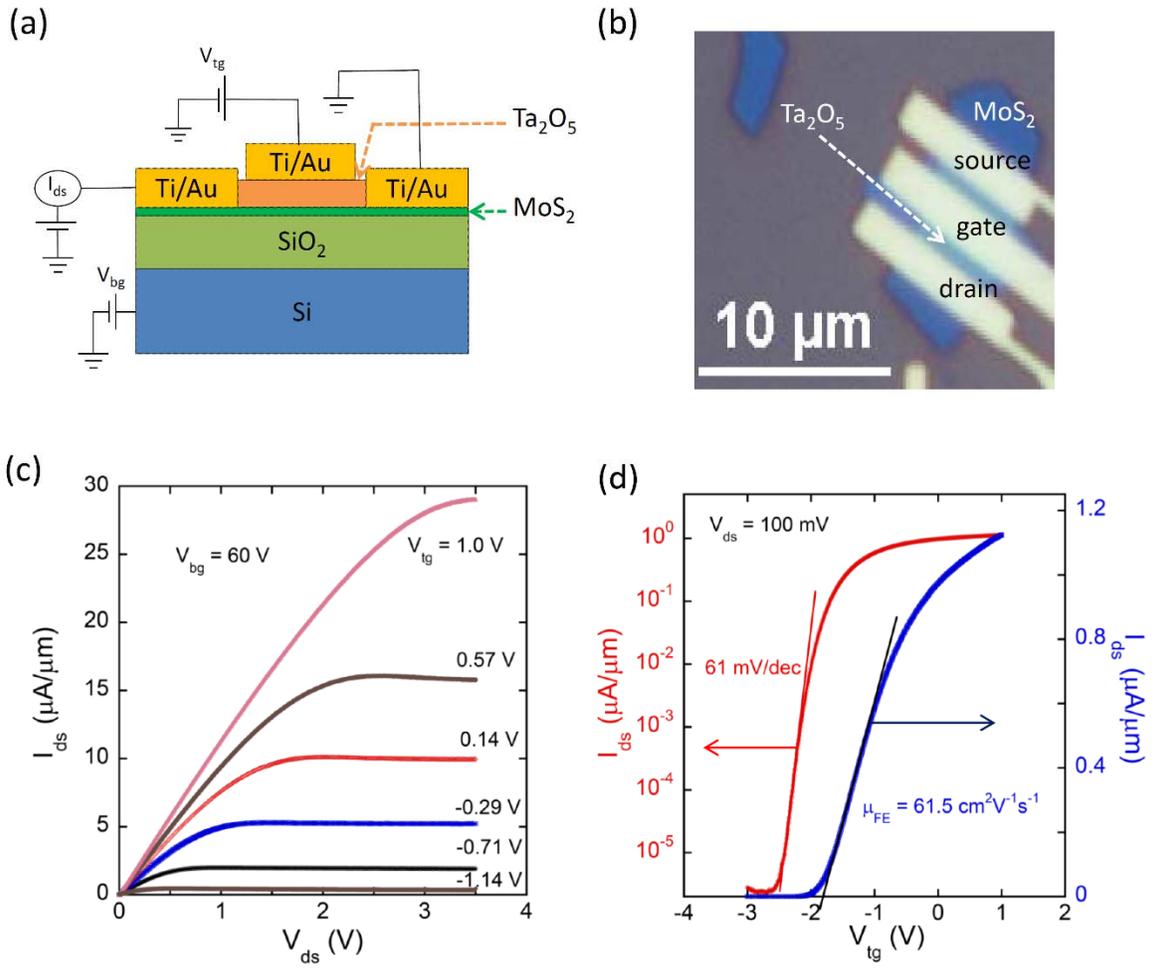

Figure 4